\documentclass[12pt]{iopart}

\usepackage{amstext}
\usepackage{iopams}  
\usepackage[sort&compress,comma]{natbib}
\usepackage{siunitx}
\usepackage{booktabs}
\usepackage{color, graphicx, amsmath,amssymb,import }
\DeclareGraphicsExtensions{.png,.pdf,.svg}
\begin{document}

\title[%
    %Decomposing MMG and EMG
    %A paradigm shift in decomposing motor units
    %Decomposition of muscle-induced bioelectromagnetic fields
     HD-MMG is superior to HD-sEMG for motor unit decomposition]
{%
    %Multichannel magnetomyography provides robust separability of motor unit action potentials: a simulation study
    %A paradigm shift in decomposing motor units: In silico trials show superiority of magnetomyography over electromyography
    High-density magnetomyography is superior to high-desnity surface electromyography for motor unit decomposition: a simulation study
}

\author{Thomas Klotz$^{1,*}$, Lena Lehmann$^{1,2,*}$, Francesco Negro$^{3,*}$ and Oliver Röhrle$^{1,2}$}

\address{$^1$ Institute for Modelling and Simulation of Biomechanical Systems, University of Stuttgart, Pfaffenwaldring 5a, 70569 Stuttgart, Germany\\[2mm]
$^2$ Stuttgart Centre for Simulation Science (SC SimTech), Pfaffenwaldring 5a, 70569 Stuttgart, Germany\\[2mm]
$^3$ Department of Clinical and Experimental Sciences, Universi\`{a} degli Studi di Brescia, Viale Europa 11, 25123 Brescia, Italy}
\ead{francesco.negro@unibs.it, thomas.klotz@imsb.uni-stuttgart.de}
\vspace{10pt}
\begin{indented}
    \item[] $^*$ Those authors have equally contributed
    %\item[]April 2022
\end{indented}

\begin{abstract}
    \textit{Objective:}
    Studying motor units (MUs) is essential for understanding motor control, the detection of neuromuscular disorders and the control of human-machine interfaces.
    Individual motor unit firings are currently identified \textit{in vivo} by decomposing electromyographic (EMG) signals.
    Due to our body’s properties and anatomy, individual motor units can only be separated to a limited extent with surface EMG.
    Unlike electrical signals, magnetic fields do not interact with human tissues.
    This physical property and the emerging technology of quantum sensors make magnetomyography (MMG) a highly promising methodology.
    However, the full potential of MMG to study neuromuscular physiology has not yet been explored.\\
    \textit{Approach:}
    In this work, we perform \textit{in silico} trials that combine a biophysical model of EMG and MMG with state-of-the-art algorithms for the decomposition of motor units.
    This allows the prediction of an upper-bound for the motor unit decomposition accuracy.\\
    \textit{Main results:}
    It is shown that non-invasive high-density MMG data is superior over comparable high-density surface EMG data for the robust identification of the discharge patterns of individual motor units.
    Decomposing MMG instead of EMG increased the number of identifiable motor units by \SI{76}{\%}.
    Notably, MMG exhibits a less pronounced bias to detect superficial motor units.\\  
    \textit{Significance:}
    The presented simulations provide insights into methods to study the neuromuscular system non-invasively and \textit{in vivo} that would not be easily feasible by other means. 
    Hence, this study provides guidance for the development of novel biomedical technologies.
\end{abstract}

\vspace{2pc}
\submitto{\JNE}
\maketitle
\ioptwocol

% ----------------------------------------------- % 
% ----------------------------------------------- % 
\section{Introduction}
% ----------------------------------------------- % 
% ----------------------------------------------- % 
The robust identification of the discharge times of individual motor units (MUs) during voluntary contractions is essential for studying human motion \citep[cf.][]{Heckman2004,Heckman2012} or to drive human-machine interfaces \citep[e.g.,][]{Farina2012,Holobar2021}.
A motor unit consists of a motor neuron and all the muscle fibres, that it innervates.
The neuromuscular junction shows a characteristic one-by-one transmission, i.e.,
each motor neuron discharge (synchronously) triggers an action potential in all muscle fibres belonging to the respective motor unit \citep{Enoka2008}.
Thus, a muscle can be considered a natural amplifier of motor neuron activity \citep[e.g.,][]{Farina2014, Roehrle2019}.
This can be exploited to reconstruct the activity of individual motor neurons by decomposing muscle signals, i.e., mainly electromyography (EMG), into the contribution of individual motor units.\\
EMG is caused by the muscle fibre action potentials.
In detail, EMG records the resultant electric potential field due to ionic currents that cross the muscle fibre membrane.
For many years, \textit{in vivo} studies of motor units have been almost exclusively possible using invasive EMG \citep[e.g.,][]{McGill2005,Merletti2009}.
This is mainly due to the high spatial selectivity of intramuscular EMG recordings.
In the last decade, the combination of high-density surface EMG \citep[e.g.,][]{Blok2002,Merletti2016techniques} and sophisticated blind source separation algorithms \citep[e.g.,][]{Holobar2007,Chen2015,Negro2016} has enabled non-invasive motor unit decomposition.
However, due to the increased distance between the electrodes and the sources, surface EMG signals exhibit a spatial low-pass filter effect, which depends on the anatomy and the properties of the body \citep[][]{Roeleveld1997,Lowery2002,Farina2002}.
This inherently limits the number of motor units that can be reliably decomposed using high-density surface EMG.\\
In fact, the electrical activity of the muscles also causes low-amplitude currents, which induce tiny magnetic fields, i.e., in the picotesla range.
Measuring the magnetic field generated by the activation of muscle fibres is known as magnetomyography (MMG) \citep[][]{Cohen1972}.
EMG and MMG originate from the same biophysical phenomenon and contain similar information regarding the state of the muscle as well as the neural drive to the muscle.
However, the magnetic permeability of human tissues is (almost) the same as in free space \citep[][]{Malmivuo1995,Oschman2002}.
Unlike EMG, the propagation of magnetic fields is therefore hardly influenced by the properties and anatomy of the body.
Hence, non-invasive MMG measurements have the potential to overcome some of the previously described limitations of surface EMG.\\
Although MMG was already introduced in the 1970s by \citet{Cohen1972}, several challenges have limited its practical use.
Most importantly, the amplitude of the muscle-induced magnetic field is about one million times lower than the Earth's magnetic field.
Hence, MMG requires highly sensitive magnetometers and appropriate shielding from electromagnetic noise.
In recent years, the technical limitations of magnetometers have been solved to such an extent \citep[cf. e.g.,][]{Boto2017,Murzin2020,Zuo2020} that it is now possible to explore the use of MMG for biomedical applications \citep[e.g.,][]{Broser2018,Broser2021,Llinas2020}.
Therefore, it is essential to support empirical observations from experiments with a solid theoretical understanding of MMG signals \citep[e.g.,][]{Klotz2022}.\\
The aim of this study is to explore the potential of using high-density MMG for non-invasive motor unit decomposition.
For this purpose, we have developed a motor unit decomposition \textit{in silico} trial framework.
This allows to systematically compare the results obtained from the decomposition of high-density MMG and high-density surface EMG, which is currently the gold standard.
In an \textit{in silico} environment, it is possible to integrate the full knowledge of the forward model into the spike train estimation procedure.
Hence, the obtained results are an upper limit for the achievable decomposition accuracy.

% ----------------------------------------------- % 
% ----------------------------------------------- % 
\section{Methods}
% ----------------------------------------------- % 
% ----------------------------------------------- % 

\begin{figure*}[ht!]
    \center
    \includegraphics[width=1.0\textwidth]{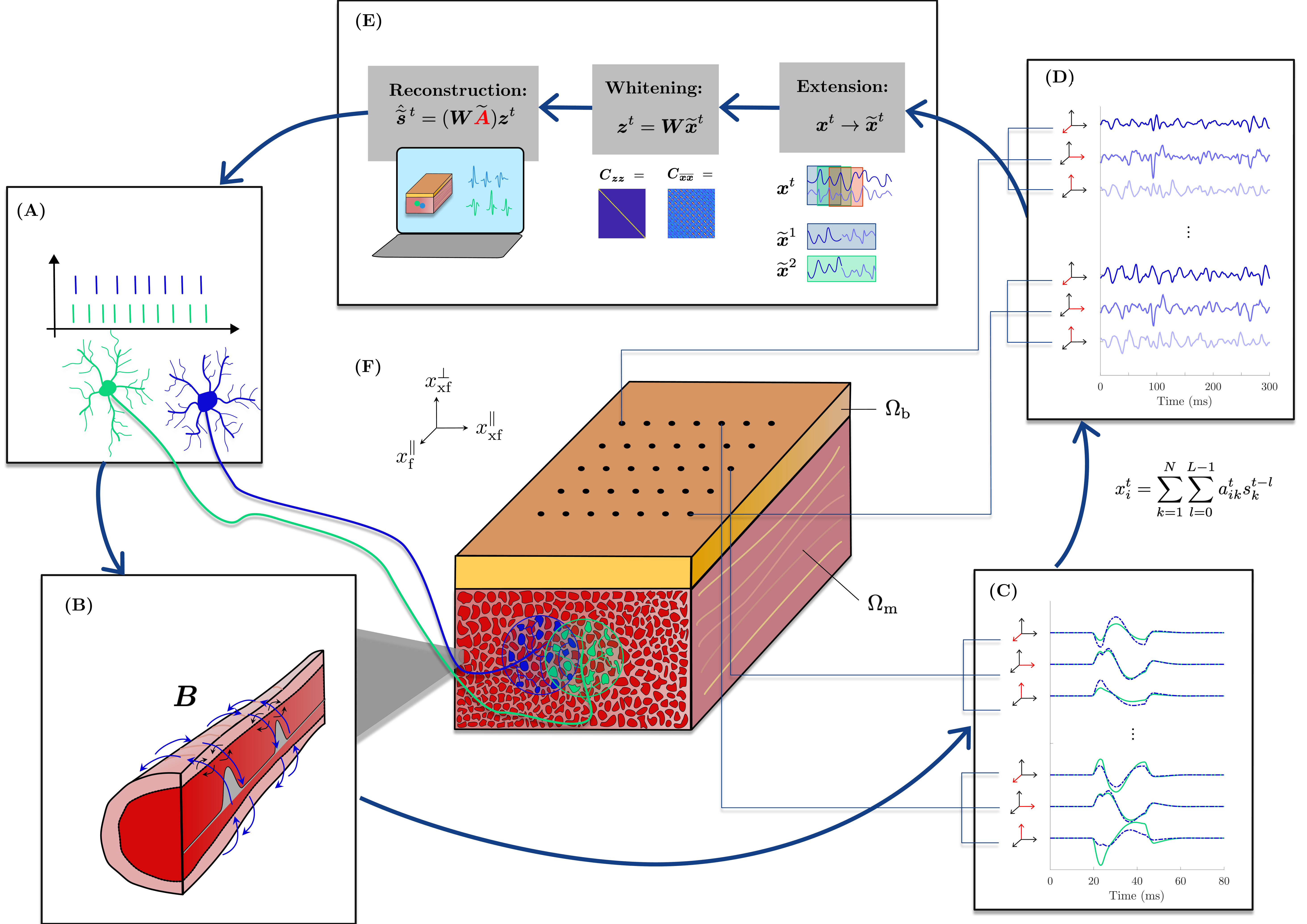}
    \caption[]{
        A-D: Schematic illustration of the generative model of MMG during voluntary contractions.
        (A) Discharge patterns of spinal motor neurons.
        (B) Every motor neuron discharge triggers an action potentials in the innervated muscle fibres.
        (C)  The spatio-temporal summation of the action potentials yields for each motor unit a characteristic motor unit magnetic field (MUMF).
        (D) The simultaneous activity of multiple motor units mixes the MUMFs according to the firing instances of the motor neurons.
        (E) Algorithm to predict upper-bound accuracy estimates of motor unit decompositions.
        (F) Illustration of the simulated tissue geometry.
    }\label{fig:graphical_abstract}
\end{figure*}

A graphical illustration of the proposed methodology is shown in Figure~\ref{fig:graphical_abstract}.
The detailed presentation of the whole methodology is structured as follows:
first, in Section~\ref{sec:methods_model} the computational methods to simulate EMG and MMG are described.
Further, Section~\ref{sec:methods_upper_bound_estimates} describes the motor unit decomposition as part of the \textit{in silico} trial framework.

% ----------------------------------------------- % 
\subsection{Modeling EMG and MMG}\label{sec:methods_model}
% ----------------------------------------------- %

% ----------------------------------------------- % 
\subsubsection{Signal Model}\label{sec:methods_mixing_model}
% ----------------------------------------------- %
EMG and MMG of voluntary contractions can be modelled as a linear convolutive mixture (cf. Figure~\ref{fig:graphical_abstract}~A-D).
For a multi-channel recording system, the EMG or MMG signal at the $i$th channel ($i=1,...,M$) is given by
\begin{equation}\label{eq:mixing_model}
    x_{i}^{t} \ =\ \sum _{k=1}^{N}\sum _{l=0}^{L-1} a_{ik}^{l} s_{k}^{t-l}  \ . 
\end{equation}
Therein, the superscript $t$ refers to a discrete time instance. 
Consequently, EMG or MMG can be understood as the superposition of all (active) motor units $k$ ($k=1,...,N$), whereby $N$ is the total number of motor units.
The contribution of each motor unit $k$ is determined by the motor neuron discharge times, i.e., a binary spike train $s_k^t$, 
which is convolved by a corresponding motor unit response $\boldsymbol{a}_{ik} = [a_{ik}^0, ..., a_{ik}^{L-1}]^T$.
Thereby, $L$ denotes the number of time samples of the (discrete) motor unit response.
A motor unit response is defined as the signal generated by the (synchronised) activity of all muscle fibres innervated by the same motor neuron.
In the following, the terms motor unit electric potential (MUEP) and motor unit magnetic field (MUMF) are used to distinguish motor unit responses observed by means of EMG and MMG, respectively.
Note that for high-density EMG each channel represents one sampling point.
For a high-density MMG signal, each sampling point is associated with three channels, i.e., one channel for each component of the magnetic field vector (cf. Figure~\ref{fig:graphical_abstract}~C-D).

% ----------------------------------------------- % 
\subsubsection{The bioelectromagnetic skeletal muscle model}\label{sec:multi-domain}
% ----------------------------------------------- %
MUEPs and MUMFs depend on the properties of the muscle fibres, a motor unit's fibre load, the geometry of the motor unit territories as well as the electric properties of the body.
To simulate MUEPs and MUMFs for a population of virtual muscles, a biophysical model is required.
Within this work, we use the multi-scale simulation framework proposed by \citet{Klotz2020,Klotz2022} to simulate a library of MUEPs and MUMFs.
In short, we assume quasi-static conditions to simulate muscle-induced bioelectromagnetic fields \citep{Griffiths2013,Malmivuo1995}.
This decouples the phenomena and allows to first solve the electric field equations before solving the magnetic field equations.\\
In the model, a skeletal muscle composed of muscle fibres of different motor units and connective tissue is homogenised. That is, at every given material point there exist one extracellular domain and one intracellular domain for each motor unit. 
Accordingly, the total current density can be defined as
\begin{equation}
    \boldsymbol{j} \ = \ \boldsymbol{j}_\mathrm{e} \, + \, \sum_{k=1}^N f_\mathrm{r}^k \boldsymbol{j}_\mathrm{i}^k \ , 
\end{equation}
where $f_\mathrm{r}^k$ is the (local) muscle fibre density of motor unit $k$ (with $\sum_{k=1}^N f_\mathrm{r}^k = 1$).
Further, Maxwell's equations require that the current density is conserved. 
This holds, if one assumes that the extracellular space and the intracellular spaces are coupled through the transmembrane current densities $I_\mathrm{m}^k$, i.e.,
\begin{subequations}
    \begin{align}
        \operatorname{div} \boldsymbol{j}_\mathrm{e} \ &= \ - \sum_{k=1}^N f_\mathrm{r}^k A_\mathrm{m}^k I_\mathrm{m}^k \ , \\
        \operatorname{div} \boldsymbol{j}_\mathrm{i}^k \ &= \ A_\mathrm{m}^k I_\mathrm{m}^k \ .       
    \end{align}
\end{subequations}
Therein, $A_\mathrm{m}^k$ is the surface-to-volume ratio of a muscle fibre associated with motor unit $k$.
The transmembrance currents are computed with a microscopic electric circuit model \citep[][]{Hodgkin1952}.
Further, each domain is considered as a volume conductor with an individual electric potential and conductivity tensor. 
Mathematically, this yields for every skeletal muscle material point $\mathcal{P}\in \mathrm{\Omega_m}$ a set of coupled differential equations:
\begin{subequations}
    \begin{align}
        0 \ & = \ \operatorname{div} \left[\boldsymbol{\sigma}_\mathrm{e} \operatorname{grad} \phi_\mathrm{e} \right] \, \notag \\
            & \! \! \! \! + \, \sum_{k=1}^N  \, f_\mathrm{r}^k \operatorname{div}
        \left[\boldsymbol{\sigma}_\mathrm{i}^k \operatorname{grad} \left(V_\mathrm{m}^k + \phi_\mathrm{e} \right) \right] \ , \label{eq:extracellular_md_eqs}  \\
        \displaystyle\frac{\partial V_\mathrm{m}^k}{\partial t} \ & = \
        \frac{1}{C_\mathrm{m}^k A_\mathrm{m}^k} \Big( \operatorname{div} \, \left[\boldsymbol{\sigma}_\mathrm{i}^k \operatorname{grad} \left(V_\mathrm{m}^k  + \phi_\mathrm{e} \right) \right] & \notag  \\
           & \! \! \! \! \! \! \! \! - \, A_\mathrm{m}^k I_\mathrm{ion}^k(\boldsymbol{y}^k,V_\mathrm{m}^k,I_\mathrm{s}^k) \Big) \ , \ k=1,...,N , \label{eq:intracellular_md_eqs} \\
        \dot{\boldsymbol{y}}^k \ & = \ \boldsymbol{g}^k(\boldsymbol{y}^k,V_\mathrm{m}^k, I_\mathrm{s}^k) \ , \ \ \ \, k=1,...,N  . \label{eq:intracellular_md_eqs_odes}
    \end{align}
\end{subequations}
Thus, the resulting multi-scale skeletal muscle model describes the evolution of the transmembrane potential $V_\mathrm{m}^k$ of the (homogenised) muscle fibres of a motor unit $k$ and the extracellular potential $\phi_\mathrm{e}$ in response to the motor neuron firings. 
That is, the post-synaptic current pulses $I_\mathrm{s}^k$ at the neruomuscular junction.
The active response of the muscle fibre membranes is simulated with an electric circuit model that takes into account a sodium conductance, a potassium conductance and a leakage conductance \citep{Hodgkin1952}.
Mathematically, the dynamic ionic permeability $I_\mathrm{ion}^k$ of the muscle fibre membranes is described by a set of ordinary differential equations summarised by Equation~\ref{eq:intracellular_md_eqs_odes}.
Further, $C_\mathrm{m}^k$ is the capacitance per unit area for a patch of the membrane of a muscle fibre associated with motor unit $k$.\\
To simulate the muscle-induced electric potential field in the entire body, the electrophysiological skeletal muscle model can be coupled to a volume conductor model, e.g., describing the influence of the subcutaneous fat or the skin.
This yields a generalised Laplace equation, i.e.,
\begin{equation}\label{eq:body_pot}
    \operatorname{div} \left[ \boldsymbol{\sigma}_\mathrm{b} \operatorname{grad} \phi_\mathrm{b} \right] \ = \ 0 \ \ \text{in} \ \mathrm{\Omega_b} \ .
\end{equation}
Therein, $\phi_\mathrm{b}$ and $\boldsymbol{\sigma}_\mathrm{b}$ denote the electric potential and the conductivity tensor in the body region $\mathrm{\Omega_b}$, respectively.\\
The solution of the model requires suitable boundary conditions. That is, no current leaves the body at its boundary. To couple the regions, it is assumed that at the muscle-body interface the potentials $\phi_\mathrm{e}$ and $\phi_\mathrm{b}$ are continuous. Further, it is required that any current leaving the extracellular space must enter the body region. Finally, it is assumed that no current can leave the intracellular domains at the boundary of the muscle. This is equivalent to a sealed ending in a cable model. In summary:
\begin{subequations}
    \begin{align}
        (\boldsymbol{\sigma}_\mathrm{b} \operatorname{grad}\phi_\mathrm{b})  \cdot \boldsymbol{n}_\mathrm{b} \ = \ 0 \ &\text{on} \ \mathrm{\Gamma}_\mathrm{b} \ , \\
        (\boldsymbol{\sigma}_\mathrm{e} \operatorname{grad}\phi_\mathrm{e})  \cdot \boldsymbol{n}_\mathrm{m} \ = \ 0 \ &\text{on} \ \mathrm{\Gamma_m} \setminus \mathrm{\Gamma_b} \ , \\
        \phi_\mathrm{e} - \phi_\mathrm{b} \ = \  0  \ &\text{on}  \ \mathrm{\Gamma_m} \cap \mathrm{\Gamma_b}  \ , \\
        (\boldsymbol{\sigma}_\mathrm{e}\operatorname{grad}\phi_\mathrm{e}  -  \boldsymbol{\sigma}_\mathrm{b}\operatorname{grad}\phi_\mathrm{e})&\cdot \boldsymbol{n}_\mathrm{m} \notag \\ \ = \ 0 \ &\text{on} \ \mathrm{\Gamma_m} \cap \mathrm{\Gamma_b}  \ , \\
        (\boldsymbol{\sigma}_\mathrm{i}^k\operatorname{grad}V_\mathrm{m}^k  +  \boldsymbol{\sigma}_\mathrm{e}\operatorname{grad}\phi_\mathrm{e})&\cdot \boldsymbol{n}_\mathrm{m} \notag \\ \ = \ 0 \ &\text{on} \ \mathrm{\Gamma}_\mathrm{m}  \ .
    \end{align}
\end{subequations}
Therein, $\boldsymbol{n}_\mathrm{b}$ and $\boldsymbol{n}_\mathrm{m}$ are unit outward vectors of the body surface $\mathrm{\Gamma}_\mathrm{b}$ and the muscle surface $\mathrm{\Gamma}_\mathrm{m}$, respectively.\\
With the solution of the electric field problem at hand we can solve the magnetic field problem.
In detail, the muscle-induced magnetic field at observation point $\boldsymbol{r}$ is determined by the Biot-Savart law \citep{Griffiths2013}, i.e., 
\begin{equation}
    \boldsymbol{B}(\boldsymbol{r}) \ = \ \frac{\mu_0}{4\pi} \iiint_{V} \frac{\boldsymbol{j} \times \boldsymbol{r}'}{|\boldsymbol{r}'|^3} \, \text{d}V \ .
\end{equation}
Therein, $\mu_0$ is the vacuum permeability, $\boldsymbol{j}$ is the total current density and $\boldsymbol{r}'$ is a vector from a material point to $\boldsymbol{r}$. 
Further, the volume integral is evaluated for the whole simulated tissue sample, i.e., $\Omega_\mathrm{m} \cup \Omega_\mathrm{b}$.
Recalling that it is assumed that the total current density is purely conductive (cf. \citet{Klotz2022}), the total current density is related to the electric potential fields via Ohm's law:
\begin{equation}\label{eq:current_muscle}
    \begin{aligned}
    \boldsymbol{j} \ = \ &\boldsymbol{\sigma}_\mathrm{e} \operatorname{grad} \phi_\mathrm{e}                                                               \\
        & + \, \sum_{k=1}^N  f_\mathrm{r}^k \boldsymbol{\sigma}_\mathrm{i}^k \operatorname{grad} (V_\mathrm{m}^k + \phi_\mathrm{e}) \ \text{in} \ \mathrm{\Omega_m} \ .
    \end{aligned}
\end{equation}
Analogous, the current density in the body region is given by
\begin{equation}\label{eq:current_body}
   \boldsymbol{j} \ = \ \boldsymbol{\sigma}_\mathrm{b} \operatorname{grad} \phi_\mathrm{b} \ \text{in} \ \mathrm{\Omega_b} \ .
\end{equation}
Note that the presented biophysical model can only be solved numerically. The utilised discretisation scheme (finite differences) is adopted from \citet{Klotz2020,Klotz2022}.
The resultant computational model is publicly available\footnote[7]{https://bitbucket.org/klotz{\_}t/multi{\_}domain{\_}fd{\_}code}.
A summary of all model parameters is given in Table~\ref{tab:parameters}.

\begin{table*}[ht!]
    \begin{tabular}{ l l l l}
        Parameter                               & Symbol                         & Value                        & Reference                              \\
        \toprule
        Longitudinal intracellular conductivity & $\sigma_\mathrm{i}^\mathrm{l}$ & \SI{8.93}{\milli\siemens\per\centi\meter}   & \cite{Bryant1969}                      \\  %\midrule
        Transversal intracellular conductivity  & $\sigma_\mathrm{i}^\mathrm{t}$ & \SI{0.0}{\milli\siemens\per\centi\meter}    & \cite{Klotz2020}                   \\  % \midrule
        Longitudinal extracellular conductivity & $\sigma_\mathrm{e}^\mathrm{l}$ & \SI{6.7}{\milli\siemens\per\centi\meter}    & \cite{Rush1963}                        \\  % \midrule
        Transversal extracellular conductivity  & $\sigma_\mathrm{e}^\mathrm{t}$ & \SI{1.34}{\milli\siemens\per\centi\meter}   & \cite{Gielen1984}                  \\  %\midrule
        Fat conductivity                        & $\sigma_\mathrm{b}$            & \SI{0.4}{\milli\siemens\per\centi\meter}    & \cite{Rush1963}                        \\  \midrule
        Membrane capacitance                    & $C_\mathrm{m}^k$               & \SI{1}{\micro\farad\per\square\centi\meter} & \cite{Hodgkin1952}                     \\
        Surface-to-volume ratio                 & $A_\mathrm{m}^k$               & Variable        & cf. Section~\ref{sec:methods_geometry}                   \\
        Motor unit density                      & $f_\mathrm{r}^k$               & Variable                                    & cf. Section~\ref{sec:methods_geometry} \\\midrule
        Magnetic permeability                   & $\mu_0$                        &  \SI{1.257e-6}{\newton\per\square\ampere} & Physical constant \\\bottomrule
    \end{tabular}
    \caption[]{Summary of parameters for the multi-domain skeletal muscle model.}\label{tab:parameters}
\end{table*}

% ----------------------------------------------- % 
\subsubsection{Tissue geometry}\label{sec:methods_geometry}
% ----------------------------------------------- % 
The spatio-temporal pattern of MUEPs and MUMFs strongly depends on the tissue geometry, the tissue's electromagnetic properties and the characteristics of the muscle fibres.
To guarantee a controlled environment, we consider in this work a layered tissue model consisting of a cube-shaped (half) muscle, i.e., $L=\SI{80}{\milli\meter}$, $W=\SI{40}{\milli\meter}$ and $H=\SI{40}{\milli\meter}$, and a fat tissue layer on top of it (cf. Figure~\ref{fig:basic_signal_properties}B).
To test the influence of subcutaneous fat on the decomposition performance, three different fat tissue thicknesses are simulated, i.e., $\SI{0}{\milli\meter}$, $\SI{5}{\milli\meter}$ and $\SI{20}{\milli\meter}$.
To resolve the functional architecture of skeletal muscles, five different motor unit pools consisting of 150 motor units are considered.
The fibres of the different motor units are characterised by the micro-scale shape parameter $A_\mathrm{m}^k$. In detail, the surface-to-volume ratio is \SI{500}{\per\centi\meter} for the smallest motor unit and \SI{250}{\per\centi\meter} for the largest motor unit. For all other motor units the parameter $A_\mathrm{m}^k$ is linearly interpolated between the minimum and the maximum value. Notably, the utilised distribution yields muscle fibre conduction velocities ranging from \SIrange[]{3}{6}{m/s} \citep[cf.][]{Klotz2020}.
Moreover, we assume an idealised circular shape of the motor unit territories.
Hence, each motor unit territory is fully described by its centre and radius.
The motor unit territory centres are randomly distributed within the boundary of the muscle's cross-sectional area.
Further, the radius of each motor unit territory is randomised from a uniform distribution ranging from \SIrange[]{3}{5}{\milli\meter}.
It is assumed that the ratio of the innervation number of the biggest motor unit and the smallest motor unit is approximately 100.
To compute for each motor unit territory a local (continuous) fibre load, first we weight each point within the motor unit territory by a parameter $w_k$, i.e.,
\begin{equation}
    w_k \ = \ \exp\left[ \ln(100) \, \frac{(k-1)}{(N-1)} \right] \, + \, 1 \ .
\end{equation}
Next, the motor unit volume fractions $f_\mathrm{r}^k$ are calculated for each skeletal muscle material point:
\begin{equation}
    f_\mathrm{r}^k \ = \ w_k \, / \, \sum_{m=1}^N w_m \ , \ k=1,...,N \ .
\end{equation}
In a last step, the motor unit territories are sorted such that the integral of the parameter $f_\mathrm{r}^k$ over the whole muscle region increases with the motor unit index $k$.
The neuromuscular junctions are randomly distributed between \SIrange[]{10}{20}{\milli\meter} in the muscle fibre direction.

% ----------------------------------------------- % 
\subsubsection{Virtual detection system}\label{sec:methods_recording_system}
% ----------------------------------------------- %
The computational model predicts for each time step and each grid point the electric potential in each domain and the magnetic field at the specified sampling points (cf. Section~\ref{sec:multi-domain}).
Here, we assume an ideal measurement system not interfering with the physical fields and point-like MMG sensors / EMG electrodes.
To obtain high-density EMG or MMG signals we consider a $10\times7$ array of equidistantly distributed sampling points (i.e., electrodes or magnetometers).
The detection system is located in the middle between the innervation zone and the boundary of the muscle (cf. Figure~\ref{fig:basic_signal_properties}B).
Thereby, the EMG electrodes are located on the top of the tissue sample and MMG sensors are placed in a plane \SI{1}{\milli\meter} above the tissue sample. 
The spacing between the sampling points is \SI{5}{\milli\meter}.
The sampling frequency is \SI{2000}{Hz}.
To simulate high-density surface EMG at each sampling point the body potential $\phi_\mathrm{b}$ is recorded.
In the case where subcutaneous fat is neglected the extracellular potential $\phi_\mathrm{e}$ is measured.
For the virtual high-density MMG recordings, we consider a vector magnetometer system, which simultaneously measures at each sampling point all three components of the magnetic field vector.
Hence, while the high-density EMG records $70$ variables, the high-density MMG measures $210$ variables.

% ----------------------------------------------- % 
\subsubsection{Motor neuron model}\label{sec:methods_voluntary_drive}
% ----------------------------------------------- %
Muscle force can be modulated by motor unit recruitment and rate coding \citep{Kandel2000,Heckman2004,Heckman2012}.
Here, we use a phenomenological approach that integrates basic physiological knowledge to obtain motor unit spike trains for three different isometric contraction levels, i.e., low intensity, medium intensity, and high intensity.
First, it is assumed that a motor unit's recruitment threshold increases with the motor unit's size \citep{Henneman1965}.
Hence, the number of active motor units increases with the contraction intensity (cf. Table~\ref{tab:MU_firings}).
Further, it is assumed that the smallest motor unit exhibits the highest firing frequency and that the peak firing rate increases with the contraction intensity (cf. Table~\ref{tab:MU_firings}).
The firing rate of the largest active motor unit is always set to \SI{8}{Hz} \citep{Duchateau2011}.
Finally, the firing frequencies for all other motor units are uniformly distributed between the minimum firing rate (\SI{8}{Hz}) and the peak firing rate.
Note that for each contraction level the firing rates decrease with the size of the motor units.
Spike trains are obtained by adding a random jitter of \SI{\pm 10}{\percent} of the mean inter-spike interval to the base firings \citep{Clamann1969,Matthews1996}.\\ 

\begin{table}[h!]
    \begin{tabular}{ l  c  c}
        \toprule
        Intensity & Recruited MUs & Peak firing rate \\ \toprule
        Low       & 60            & \SI{15}{Hz}      \\
        Medium    & 100           & \SI{20}{Hz}      \\
        High      & 150           & \SI{25}{Hz}      \\ \bottomrule
    \end{tabular}
    \caption[]{Motor neuron model parameters.}\label{tab:MU_firings}
\end{table}

% ----------------------------------------------- % 
\subsection{Upper-bound accuracy estimates of motor unit decompositions}\label{sec:methods_upper_bound_estimates}
% ----------------------------------------------- % 

% ----------------------------------------------- % 
\subsubsection{Spike train estimation}\label{sec:methods_pipeline}
% ----------------------------------------------- % 
The main aim of this work is to explore the potential of using MMG to record the firings of individual motor units non-invasively and \textit{in vivo}.
For this purpose, we have developed an \textit{in silico} trial framework. This allows to systematically compare the results obtained from high-density MMG-based and high-density surface EMG-based motor unit decompositions. 
In short, the decomposition scheme utilised follows closely algorithms for blind source separation \citep[e.g.,][]{Negro2016,Holobar2007,Chen2015}.
However, by considering simulated signals, it is possible to directly integrate the full knowledge about the forward mixing model, i.e., the motor unit responses, into the spike train estimation step.
Hence, the reconstructed spike trains represent upper-bound accuracy estimates.
Further, the predicted spike trains are not affected by the choice of a specific decomposition algorithm.
The proposed method is schematically illustrated in Figure~\ref{fig:graphical_abstract}E and is implemented in MATLAB (The MathWorks, Inc., Natick, Massachusetts, United States).\\

Mathematically, motor unit decomposition requires to invert the signal model given in Equation~\eqref{eq:mixing_model}.
Therefore, one makes use of the fact that any convolutive mixture with finite impulse response filters can be transformed into an instantaneous mixture of extended sources and hence, a matrix system \citep{Negro2016}.
That is, the signal $x_i^t$, consisting of $M$ observations at time frame $t$, is expanded by $K$ time instances to yield an extended observation vector $\widetilde{\boldsymbol{x}}^t \in \mathbb{R}^{KM}$:
\begin{equation}\label{eq:signal_extension}
    \begin{aligned}
        \widetilde{\boldsymbol{x}}^t \ = \ \Big[x_1^t,x_1^{t-1},... \, ,x_1^{t-(K-1)}, \, ... \, , &             \\
        x_M^t,x_M^{t-1},... \, ,x_M^{t-(K-1)}                                                      & \Big]^T \ .
    \end{aligned}
\end{equation}
Here, we chose the extension factor $K$ equal to the length of the motor unit responses, i.e., $K=L$ (cf. Section~\ref{sec:methods_mixing_model}).
The mixing of the extended observations $\widetilde{\boldsymbol{x}}^t$ can be written as
\begin{equation}\label{eq:inst_mixture_model}
        \widetilde{\boldsymbol{x}}^t \ = \  \widetilde{\boldsymbol{A}} \ \widetilde{\boldsymbol{s}}^t \ .
\end{equation}
Therein, the extended spike train vector $\widetilde{\boldsymbol{s}}^t \in \mathbb{R}^{N( L+K-1)}$ is derived from the binary spike trains $s_k^t$, i.e., %$\boldsymbol{s}^t = [s_1^t,...,s_N^t]^T$ via
\begin{equation}\label{eq:spike_train_extension}
    \begin{aligned}
        \widetilde{\boldsymbol{s}}^t \ = \ \Big[s_1^t,s_1^{t-1},...\, ,s_1^{t-(L-1)-(K-1)},...\, , &            \\
        %& s_1^{n-(L-1)-(K-1)}, ... , \\
        s_N^t,s_N^{t-1},...,s_N^{t-(L-1)-(K-1)}\Big]                                               & ^T    \, . \\
        %& s_N^{n-(L-1)-(K-1)}\Big]^T \ .
    \end{aligned}
\end{equation}

Further, the mixing matrix of the extended system $\widetilde{\boldsymbol{A}} \in \mathbb{R}^{KM\times N( L+K-1)}$ is given by
\begin{equation}\label{eq:mixing_matrix}
   \begin{aligned}
        \displaystyle \widetilde{\boldsymbol{A}} \ = & \begin{bmatrix}   \widetilde{\boldsymbol{A}}_{11} & \widetilde{\boldsymbol{A}}_{12} & ... & \widetilde{\boldsymbol{A}}_{1N} \\
        \widetilde{\boldsymbol{A}}_{21} & \widetilde{\boldsymbol{A}}_{22} & ... & \widetilde{\boldsymbol{A}}_{2N} \\
        ... & ... & ... & ... \\
        \widetilde{\boldsymbol{A}}_{M1} & \widetilde{\boldsymbol{A}}_{M2} & ... & \widetilde{\boldsymbol{A}}_{MN} \end{bmatrix} \\
    \end{aligned} \ ,
\end{equation}
where the block matrices $\widetilde{\boldsymbol{A}}_{ij} \in \mathbb{R}^{K \times (L+K-1)}$ are defined as
\begin{equation}\label{eq:mixing_block_matrices}
   \begin{aligned}
        \displaystyle \widetilde{\boldsymbol{A}}_{ik} \ = & \left[\begin{smallmatrix}
        a_{ik}^0 & a_{ik}^1   & ... & a_{ik}^{L-1} & 0 &  ...  & 0\\
        0 & a_{ik}^0          &  ...& a_{ik}^{L-2} & a_{ik}^{L-1} & ... & 0\\
        . & .      & ... & . & . &  ... & .  \\
        . & .      & ... & . & . &  ... & . \\
        . & .    & ... & . & . &  ... & .  \\
        0 & 0  & ... & . & . & ... & a_{ik}^{L-1}
        \end{smallmatrix} \ \right] \ ,
    \end{aligned}
\end{equation}
and $a_{ik}^l$ are the motor unit responses, cf. Section~\ref{sec:methods_mixing_model}.

The signal extension is followed by a whitening transformation that uncorrelates the extended signals, i.e.,
\begin{equation}\label{eq:signal_whitening}
   \begin{aligned}
        \boldsymbol{z}^t \ = \ \boldsymbol{W} \widetilde{\boldsymbol{x}}^t \ . 
    \end{aligned}
\end{equation}
Here, we apply ZCA whitening \citep{Krizhevsky2009}, in which the whitening matrix $\boldsymbol{W}$ is determined by the eigendecomposition of the extended signal's covariance matrix $\boldsymbol{C}_{\widetilde{\boldsymbol{x}}\widetilde{\boldsymbol{x}}}$, i.e.,
\begin{equation}\label{eq:whitening_matrix}
    \boldsymbol{W} \ = \ \boldsymbol{V} \boldsymbol{D}^{-\frac{1}{2}} \boldsymbol{V}^T \ .
\end{equation}
Therein, $\boldsymbol{D}$ is a diagonal matrix, that contains, in increasing order, the eigenvalues $e_j$ ($j = 1,...,KM$) of $\boldsymbol{C}_{\widetilde{\boldsymbol{x}}\widetilde{\boldsymbol{x}}}$. Matrix $\boldsymbol{V}$ contains the corresponding eigenvectors.
To avoid numerical errors, eigenvalues that are numerically zero, i.e.,
$e_j < \varepsilon \cdot KM$ and $\varepsilon$ is the modulus of the distance from the maximum eigenvalue to the next larger floating point number, are discarded for the calculation of the whitening matrix $\boldsymbol{W}$.
Figure~\ref{fig:graphical_abstract}E exemplarily showcases that the covariance matrix of the extended and whitened observations is the identity matrix.
The generative model of the extended whitened signal is given by
\begin{equation}\label{eq:whitened_system}
   \begin{aligned}
        \boldsymbol{z}^t \ = \ \boldsymbol{W }\widetilde{\boldsymbol{A}} \ \widetilde{\boldsymbol{s}}^t \ = \ (\boldsymbol{W}\widetilde{\boldsymbol{A}}) \ \widetilde{\boldsymbol{s}}^t \ .
    \end{aligned}
\end{equation}
Therein, the mixing matrix $(\boldsymbol{W}\widetilde{\boldsymbol{A}})$ is equivalent to the the extended and whitened motor unit responses.

Motor unit decomposition can be achieved by inverting the linear system given in Equation~\eqref{eq:whitened_system}. 
Typically, this is an ill-conditioned problem.
Yet, the motor unit spike trains can be estimated by correlating the extended and whitened motor unit responses with the extended and whitened signal $\boldsymbol{z}^t$, i.e.,
\begin{equation}\label{eq:estimated_spike_trains}
    \hat{\widetilde{\boldsymbol{s}}}\,^t \ = \ (\boldsymbol{W}\widetilde{\boldsymbol{A}})\boldsymbol{z}^t \ = \ (\boldsymbol{W}\widetilde{\boldsymbol{A}})\boldsymbol{W}\widetilde{\boldsymbol{x}}^t \ .
\end{equation}
Therein, $\hat{\widetilde{\boldsymbol{s}}}\,^t$ is the reconstructed extended spike train at the discrete time instance $t$.
Importantly note that for experimentally measured signals the motor unit responses $\boldsymbol{a}_{ik}$ are unknown and therefore, must be approximated by an optimisation scheme.
In the presented \textit{in silico} trial framework we can exploit that the MUEPs and MUMFs are already known.
Thus, we can directly evaluate Equation~\eqref{eq:estimated_spike_trains} to obtain an optimal reconstruction of the motor unit spike trains. This provides an upper bound for the decomposition accuracy. 

In a last step, for each motor unit $k$ a binary spike train $\hat{s}^{\prime \, t}_k$ is obtained by applying a peak detection method to the estimated spike trains $\hat{s}^t_k$.  
Further, the k-means clustering algorithm is applied to identify (potential) false-positive firings \citep[cf.][]{Negro2016}, i.e., separating the spikes into two clusters whereby the centroids are initialised by the minimum peak and maximum peak height, respectively.

% ----------------------------------------------- % 
% ----------------------------------------------- %
\subsubsection{Data analysis}\label{sec:data_analysis}
% ----------------------------------------------- %
% ----------------------------------------------- % 
To quantify the uniqueness of the motor unit responses, i.e., MUEPs or MUMFs, the cosine similarity $S_\mathrm{cos}^{kq}$ ($k=1,...,N$ and $q=1,...,N$, with $k \neq q$) for all motor unit response pairs is computed. Therefore, the motor unit responses are aligned in time by maximising their channel-by-channel cross-correlation. The cosine similarity between two multi-channel motor unit responses is given by
\begin{equation}
    S_\mathrm{cos}^{kq} \ = \ \frac{\sum\limits_{l=0}^{L-1}\sum\limits_{i=1}^M a_{ik}^l a_{iq}^l}{\sqrt{\sum\limits_{l=0}^{L-1}\sum\limits_{i=1}^M \left(a_{ik}^l\right)^2}\sqrt{\sum\limits_{l=0}^{L-1}\sum\limits_{i=1}^M \left(a_{iq}^l\right)^2}} \ .% \frac{a \cdot b}{\|a\|\|b\|}
\end{equation}
Note that $S_\mathrm{cos}^{kq} = 0$ means that the $k$th and the $q$th motor unit responses are orthogonal and hence uncorrelated.
In contrast, $S_\mathrm{cos}^{kq} = 1$ indicates that the signals are identical.
To guarantee that each magnetic field component is weighted equally for the computation of the MUMF similarities, each MMG component is normalised with the maximal value observed at the highest contraction level.\\

To quantify the quality of a motor unit decomposition, a set of performance metrics is computed.
Therefore, the predicted spikes are classified for each motor unit $k$ into the true-positive spikes $\mathrm{TP}^k$, which are the firings that appear in the true spike train $s_k^t$ and in the predicted spike train $\hat{s}^{\prime \, t}_k$ with a maximal delay of \SI{\pm 0.5}{\milli\second}.
Further, the false-positive spikes $\mathrm{FP}^k$ represent the firings only included in the predicted spike train $\hat{s}^{\prime \, t}_k$ and the false-negative spikes $\mathrm{FN}^k$ denote the firings that are only included in the true spike train $s_k^t$.
Thus, for each motor unit $k$ ($k=1,...,N$) the rate of agreement (RoA) is computed by
\begin{equation}\label{eq:roa}
    \text{RoA}^k \ = \ \frac{|\text{TP}^k|}{|\text{TP}^k| + |\text{FP}^k| + |\text{FN}^k|} \ ,
\end{equation}
where $|\text{TP}^k|$ is the number of true-positive spikes, $|\text{FP}^k|$ is the number of false positive spikes and $|\text{FN}^k|$ is the number of false-negative spikes.
The uncertainty associated with the predicted spike trains $\hat{s}^{\prime \, t}_k$ is quantified by computing for each motor unit $k$ the silhouette coefficient
\begin{equation}\label{eq:sil}
   \text{SIL}^k \ = \ \tfrac{\sum \limits_{p \in \mathrm{TP}^k} D_{\mathrm{nospike}}^{pk}  - \sum \limits_{p \in \mathrm{TP}^k} D_{\mathrm{spike}}^{pk} }{\max\left(\sum \limits_{p \in \mathrm{TP}^k} D_{\mathrm{spike}}^{pk}  , \sum \limits_{p \in \mathrm{TP}^k} D_{\mathrm{nospike}}^{pk} \right)} \ .
\end{equation}
Therein, $D_{\mathrm{spike}}^{pk}$ is the euclidean distance between a true positive point of the estimated spike train $\hat{s}^p_k$ and the mean value of all estimated true positive spike points $\Bar{\hat{s}}_{p \in \mathrm{TP}^k}$, i.e., $D_{\mathrm{spike}}^{pk} = \| \hat{s}^p_k - \Bar{\hat{s}}_{p \in \mathrm{TP}^k} \|$.
Analogous, $D_{\mathrm{nospike}}^{pk}$ is the euclidean distance between a true positive value of the estimated spike train $s_{k,\mathrm{est}}^p$ and the mean value of all remaining estimated spike points not included in the set of the true-positive spikes $\Bar{\hat{s}}_{p \notin \mathrm{TP}^k}$, i.e., $D_{\mathrm{spike}}^{pk} = \| \hat{s}^p_k - \Bar{\hat{s}}_{p \notin \mathrm{TP}^k} \|$.
For $\text{SIL}^k > 0.9$ a motor unit is classified as reliably identifiable \citep{Negro2016}.

% ----------------------------------------------- % 
% ----------------------------------------------- % 
\section{Results}
% ----------------------------------------------- % 
% ----------------------------------------------- % 

\subsection{Basic signal properties}\label{sec:basic_sig}
\begin{figure*}[ht!]
    \centering
    \includegraphics[width=1.0\textwidth]{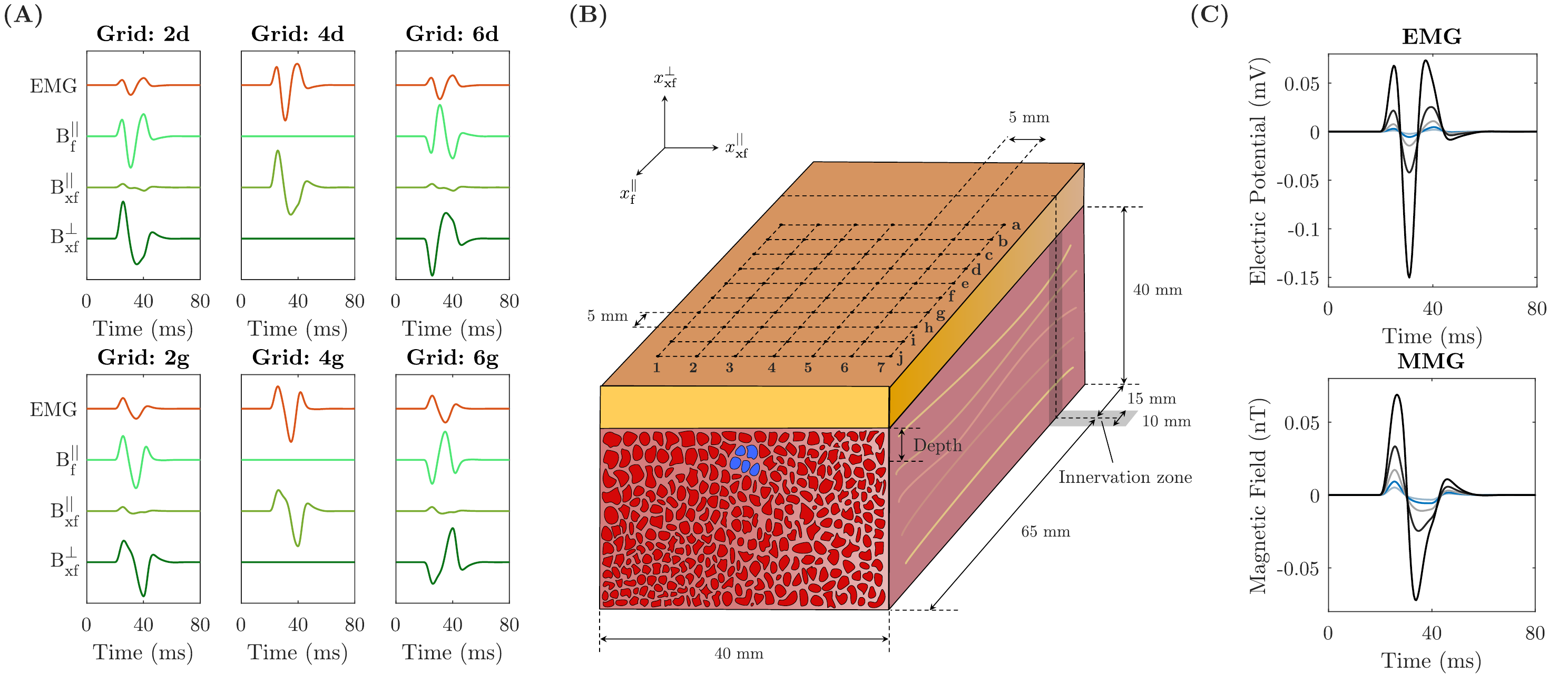}
    \caption{(A) Exemplary high-density MUEP (red) and MUMF (light green, normal green and dark green for the different componenets) at six exemplary chosen sampling points. (B) Schematic drawing of the simulated tissue geometry and the utilised high-density EMG/MMG array. (C) Time domain signal of the EMG and the $x_\mathrm{xf}^\parallel$-component of the MMG when increasing the depth of the recruited motor unit from \SIrange{3}{15}{\milli\meter} (by increments of \SI{3}{\milli\meter}) observed at one exemplary sampling point (4d).}
    \label{fig:basic_signal_properties}
\end{figure*}
First, we exemplary highlight the basic properties of MUEPs and MUMFs.
To do so, we consider a virtual muscle ($L=\SI{80}{\milli\meter}$, $W=\SI{40}{\milli\meter}$, $H=\SI{40}{\milli\meter}$) with a \SI{5}{\milli\meter} thick fat tissue layer and a motor unit that is centred with respect to the muscle's lateral axis, see Figure~\ref{fig:basic_signal_properties}B. The depth of the motor unit territory's centre is \SI{3}{\milli\meter} (independent of the fat tissue layer thickness).
It is observed that the high-density EMG or MMG signals exhibit a time delay when comparing two signals from different sampling points aligned with the muscle fibre direction.
Hence, the temporal and the spatial coordinate are linked. 
However, due to the different physical governing equations, the spatial distribution of the EMG and the MMG are fundamentally different. 
That is, the EMG amplitude is maximal directly over the source and decreases in the lateral direction.
As quasi-static magnetic fields are always perpendicular to the source currents, Figure~\ref{fig:basic_signal_properties}A shows that the spatial distribution of the MMG depends on the measured vector field component.
Like for the EMG, the MMG's $x_\mathrm{xf}^\parallel$-component is maximal over the active fibres and decreases in the lateral direction. In contrast, both the MMG's $x_\mathrm{f}^\parallel$-component and the $x_\mathrm{xf}^\perp$-component vanish directly over the active fibres and have their maxima towards the lateral direction. 
Notably, the lateral centre line represents a reflection axis, where the sign of the signal is flipped.\\
Further, Figure~\ref{fig:basic_signal_properties}C demonstrates the signal modulation of the EMG and the MMG depending on the depth of the active motor unit.
When increasing the depth of the motor unit territory's centre from \SI{3}{\milli\meter} to \SI{15}{\milli\meter}, the area under the curve drops by \SI{97.9}{\percent} for the EMG. 
For the MMG, the area under the curve decreases by \SI{93.6}{\percent}.
Moreover, increasing the depth of the motor unit territory causes a decrease of the frequency content of the motor unit responses. 
The median frequency of the signals presented in Figure~\ref{fig:basic_signal_properties}C and a motor unit depth of \SI{3}{\milli\meter} are \SI{353.7}{Hz} for the EMG and \SI{216.0}{Hz} for the MMG.
When considering the same motor unit in a depth of \SI{15}{\milli\meter} the median frequency drops by \SI{17.9}{\percent} for the EMG and by \SI{14.3}{\percent} for the MMG.

% ----------------------------------------------- % 
\subsection{Similarity of MUEPs and MUMFs}\label{res:muap_similarity}
% ----------------------------------------------- % 
\begin{figure*}[ht!]
    \centering
    \includegraphics[width=1.0\textwidth]{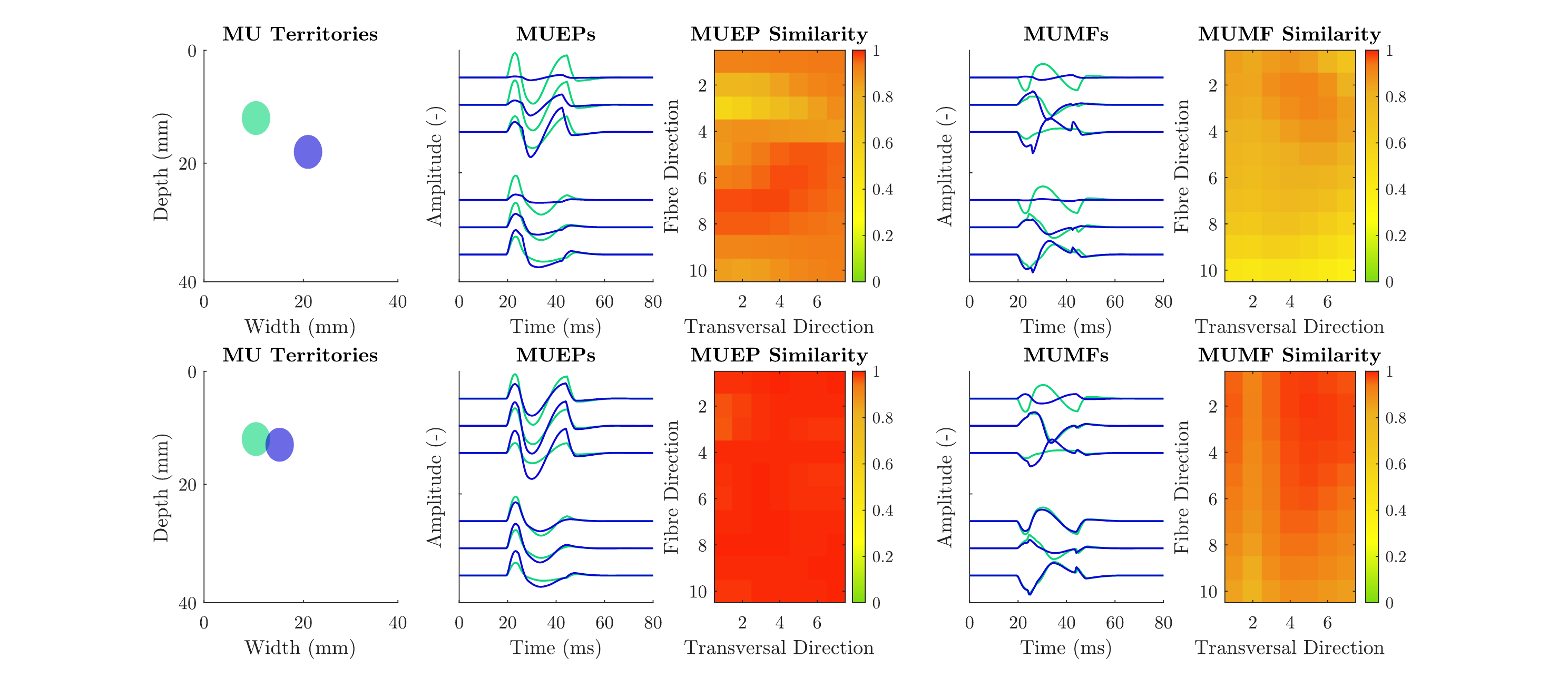}
    \caption{Similarity of the MUEPs and MUMFs from two exemplary chosen motor unit pairs. Left column: Territories of two arbitrarily chosen motor units. Second column: MUEPs on a few randomly selected channels. Third column: Channel-by-channel cosine similarity the MUEPs. A value of zero (green) indicates uncorrelated signals and a value of 1 (red) means that the signals are identical. Fourth column: MUMFs on a few randomly selected channels. Right column: Channel-by-channel cosine similarity of the MUMFs.
    }
    \label{fig:chanel_based_cos_similarity}
\end{figure*}
The unique representation of the motor unit responses in a (multi-channel) EMG or MMG recording is the basic requirement to decompose the signal into its individual components.
Surface EMG signals are low-pass filtered by the anatomy and the electric properties of the body. This inherently limits the accuracy of surface EMG-based motor unit decompositions.
To test if non-invasive MMG measurements can (partially) overcome the physical limitations of surface EMG, this section compares the similarity of MUEPs and MUMFs.
Figure~\ref{fig:chanel_based_cos_similarity} exemplary showcases for two pairs of motor units the similarity of the respective high-density MUEPs and MUEFs.
It can be observed that for two motor units with spatially distinct territories, both the MUEPs and the MUMFs, can be visually distinguished in the time domain.
In detail, the mean channel-by-channel cosine similarity is 0.90 for the MUEPs and 0.73 for the MUMFs.
Further, the minimum channel-by-channel cosine similarity is 0.54 for the MUEPs and 0.37 for the MUMFs.
However, if one considers two motor units in close proximity, the MUEPs become nearly indistinguishable.
This is also reflected in the cosine similarity, which is 0.99 across all channels and the minimum channel-by-channel cosine similarity is 0.96.
In contrast, the time domain signals of the MUMFs are still visually distinguishable for some channels.
The mean channel-by-channel cosine similarity of the two high-density MUMFs is 0.93.
Further, the minimum channel-by-channel cosine similarity is 0.79.
This example demonstrates that non-invasive MMG has advantages to separate motor units, which are hardly distinguishable using high-density surface EMG.\\

\begin{figure}[ht!]
    \centering
    \includegraphics[width=0.4\textwidth]{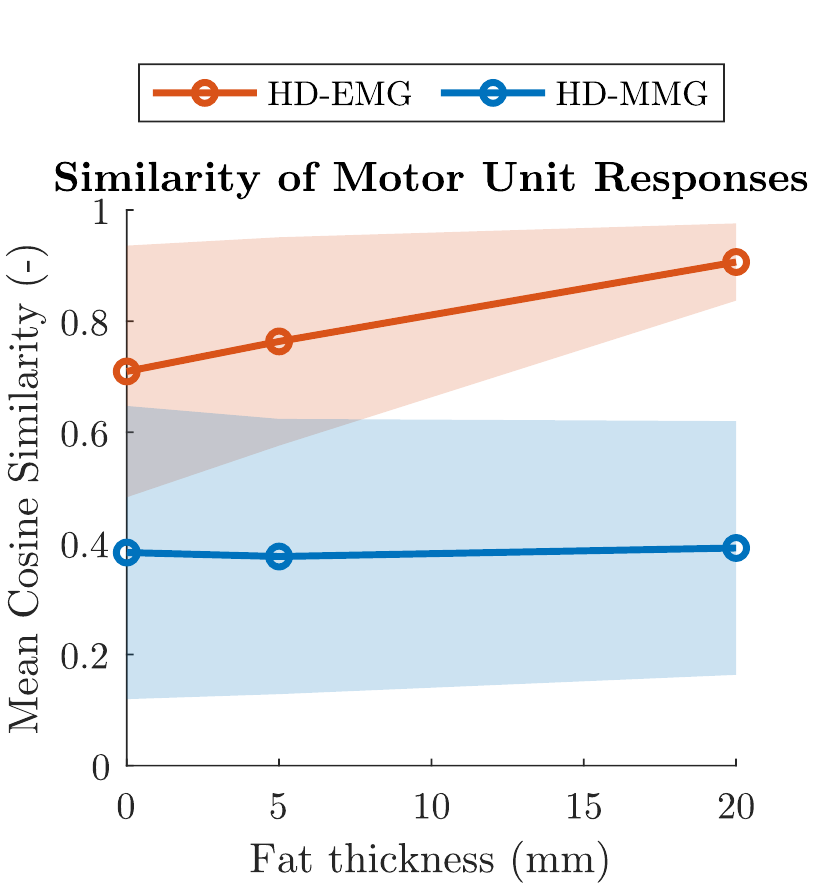}
    \caption{Mean cosine similarity of all MUEPs (red) and MUMFs (blue) in 5 motor unit pools depending on the thickness of the subcutaneous fat tissue. The circles denote the average values and the shaded areas represent the respective standard deviations.}
    \label{fig:mean_cos_similarity}
\end{figure}

The robust decomposition of a motor unit requires that the respective motor unit response is unique in comparison with all other motor unit responses.
Hence, we evaluate the similarity of the MUEPs and the MUMFs for five different motor unit populations as well as three different fat tissue thicknesses. 
Therefore, for each virtual muscle the mean cosine similarity across all MUEPs and MUMFs is computed.
Figure~\ref{fig:mean_cos_similarity} shows that the mean similarity of the MUMFs is lower than the mean similarity of the MUEPs.
For example, for a thickness of \SI{5}{\milli\meter} of the subcutaneous fat tissue, the mean cosine similarity is 0.38 for the MUMFs and 0.76 for the MUEPs.
Further, with increasing thickness of the subcutaneous fat, for the MUEPs the mean cosine similarity increases.
In contrast, there is no significant relation between the cosine similarity of the MUMFs and fat tissue thickness.
In conclusion, it can be expected that more motor units can be decomposed from the high-density MMG than from the high-density surface EMG.\\

\begin{figure}
    \centering
    \includegraphics[width=0.5\textwidth]{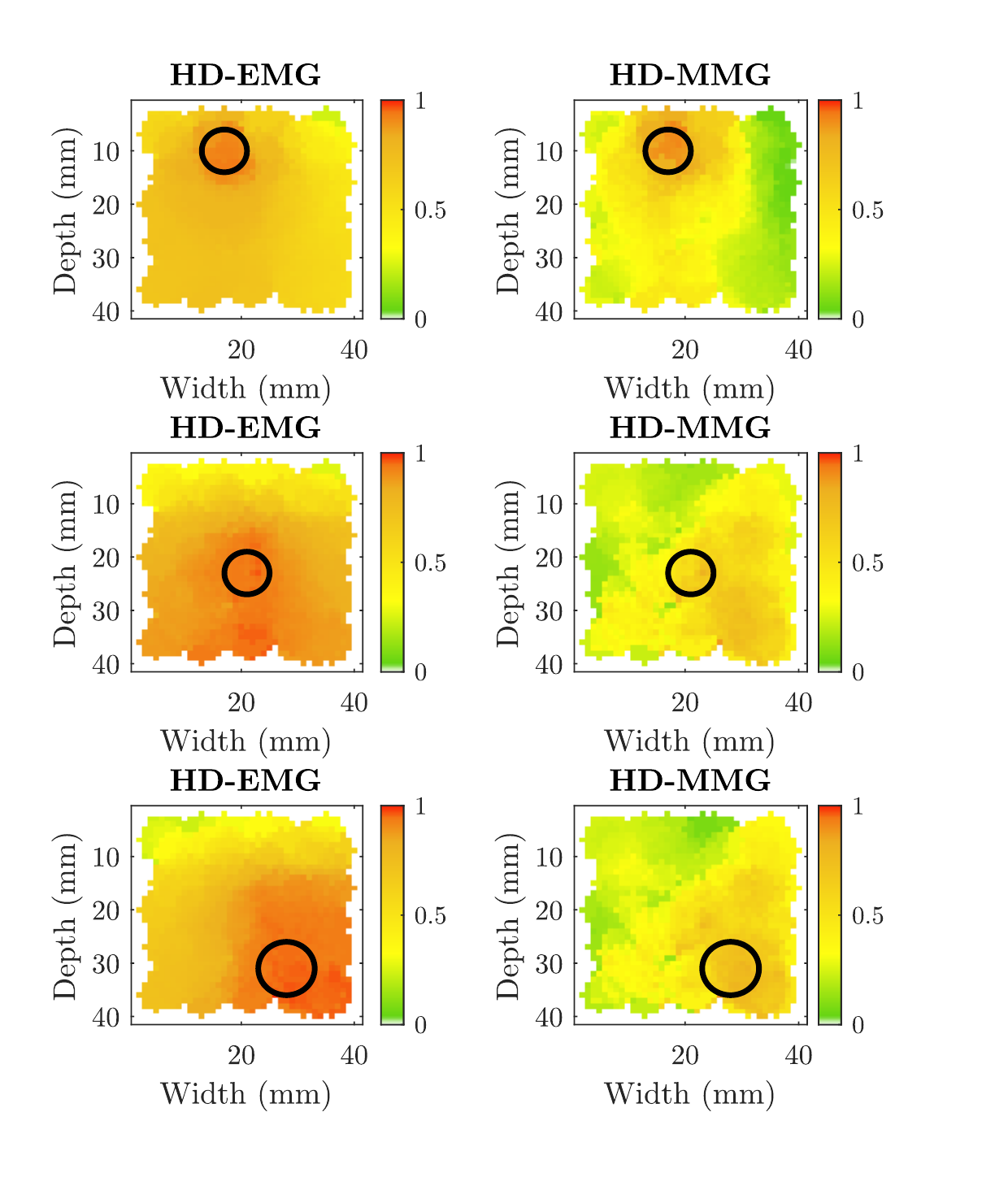}
    \caption{
        Uniqueness of three randomly selected MUEPs (left column) and MUMFs (right column) depending on the spatial position of the motor units.
        The territories of the reference motor units are visualised with black circles.
        The colour map indicates the mean cosine similarity of the reference motor unit response and all other motor unit responses from motor units that are located at the respective grid point.
    }
    \label{fig:spatial_cos_similarity}
\end{figure}

Further, Figure~\ref{fig:spatial_cos_similarity} exemplary illustrates for three motor units, i.e., superficial, centred, and deep, the influence of the spatial position of a motor unit on the uniqueness of the MUEPs and the MUMFs.
It is observed that the uniqueness of the MUEPs correlates with the depth of the motor unit territory.
That is, the fraction of the muscle's cross-sectional area with a mean cosine similarity larger than 0.9 is \SI{3.0}{\%} for the superficial motor unit, \SI{18.4}{\%} for the centred motor unit and \SI{24.5}{\%} for the deep motor unit.
For the MMG there is no considerable correlation between the depth of the motor units and the uniqueness of the MUMFs. 
In detail, the fraction of the muscle's cross-sectional area with a mean cosine similarity larger than 0.9 is always less than \SI{1}{\%}.
Thus, it is expected that the MMG decomposition will be less affected by variations in the depth of the motor units.

% ----------------------------------------------- % 
\subsection{Upper-bound decomposition accuracy for EMG and MMG}\label{res:separability}
\begin{figure}[ht!]
    \centering
    \includegraphics[width=0.5\textwidth]{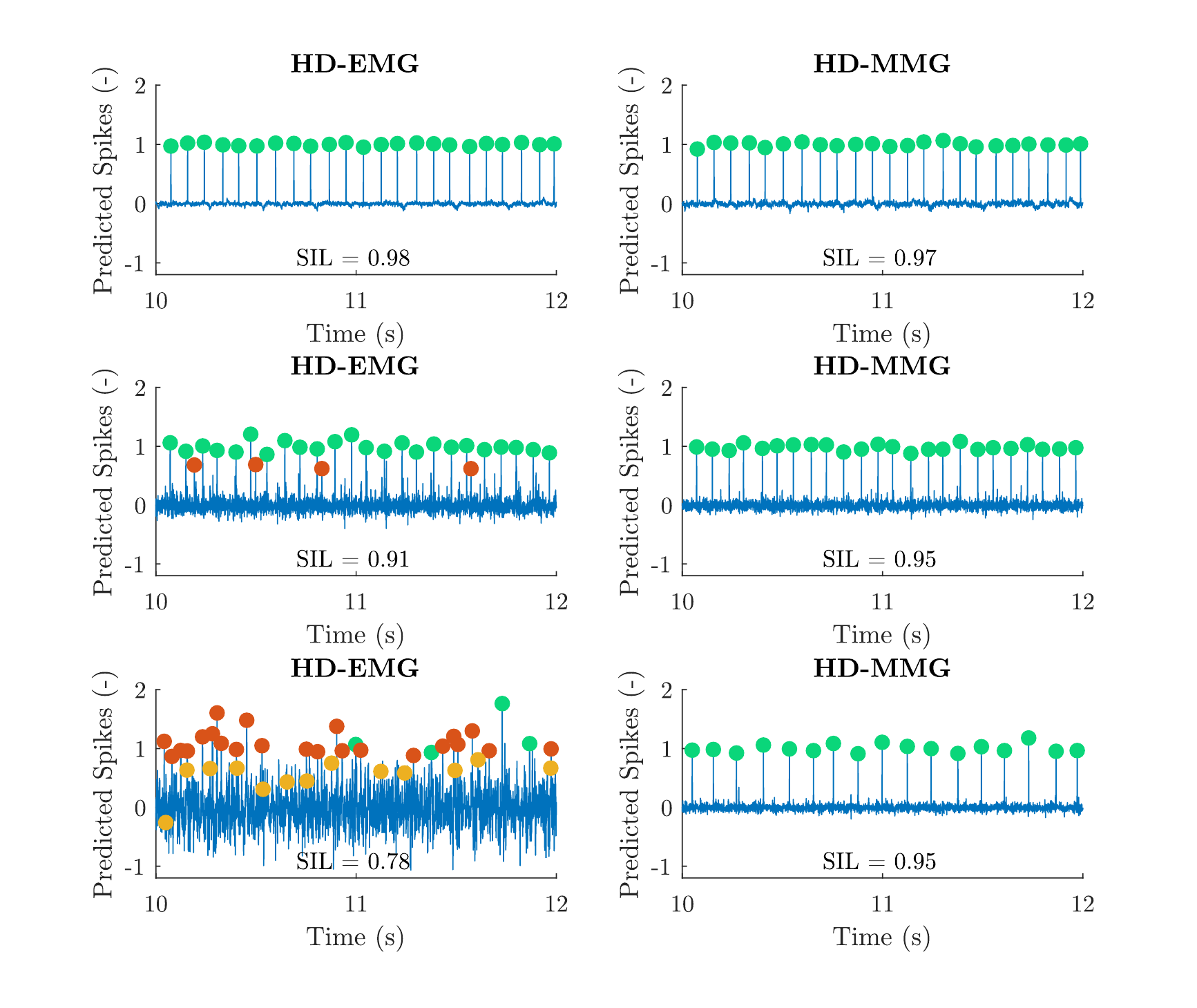}
    \caption{Spike trains estimated from HD-EMG (left column) and HD-MMG (right column). Each row represents one randomly selected motor unit. Green circles show true positive spikes, red circles show false positive spikes and orange circles show false negative spikes. The predicted spike trains are normalised with respect to the mean value of all true positive spikes.}
    \label{fig:exemplary_spike_trains}
\end{figure}

\begin{figure*}
    \centering
    \includegraphics[width=0.8\textwidth]{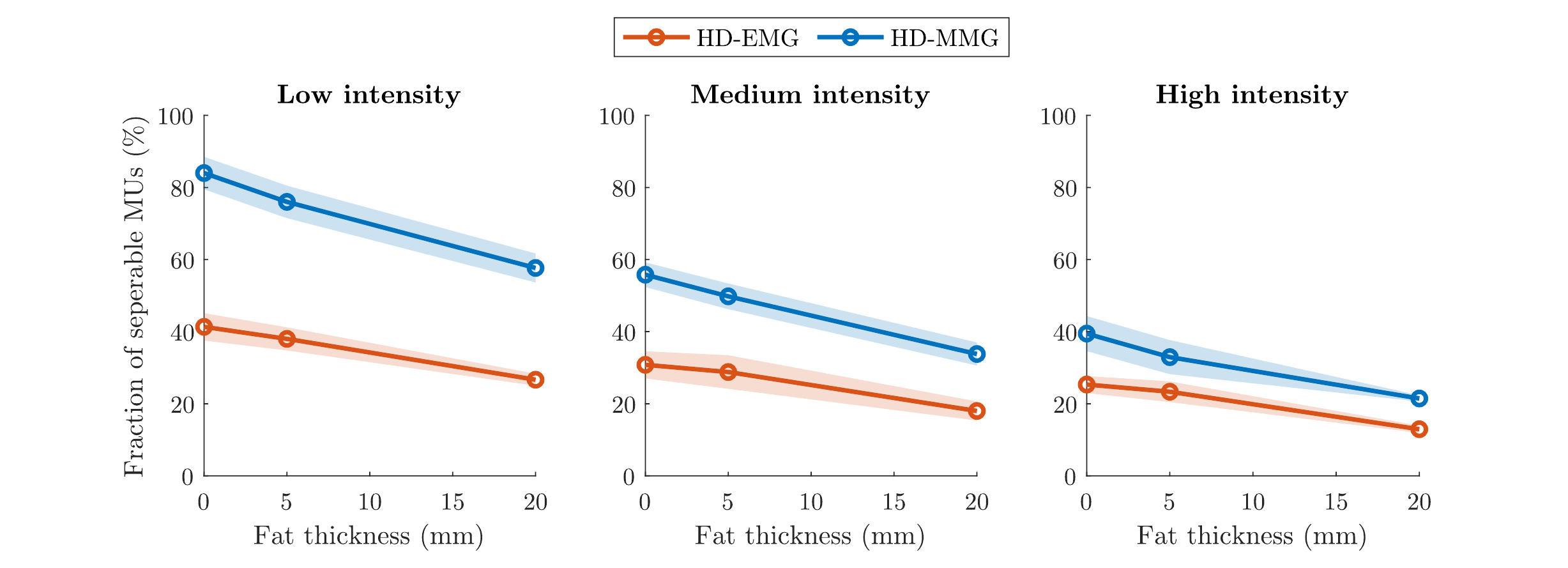}
    \caption{Fraction of motor units that are classified as reliably identifiable from high-density EMG (red) and high-density MMG (blue).
    All results are shown for three different contraction intensities (left to right) depending on the thickness of the subcutaneous fat tissue layer.
    The markers indicate mean values and the shaded areas represent the corresponding standard deviations.
    }
    \label{fig:roa_fat}
\end{figure*}

% ----------------------------------------------- % 
The motor unit decomposition results presented in this work summarise 45 independent \textit{in silico} experiments.
That is, five different motor unit pools are considered and simulated with 3 different fat tissue layers, respectively.
For those 15 virtual muscles and three different contraction intensities, i.e., low intensity, medium intensity and high intensity, \SI{30}{\second} long steady isometric contractions were simulated and analysed with the proposed methodology \citep[see][to access the replication dataset]{darus-3556_2023}.\\
Figure~\ref{fig:exemplary_spike_trains} exemplary shows for three randomly selected motor units the spike trains estimated from the high-density EMG (left column) and the high-density MMG (right column).
The first motor unit can be perfectly reconstructed both from the high-density EMG and the high-density MMG, i.e.,  the rate-of-agreement is \SI{100}{\%}.
Further, the robustness of the decomposition is reflected in the silhouette coefficients, i.e., 0.98 for the high-density EMG and 0.97 for the high-density MMG.
Also the second selected motor unit is classified as reliably identifiable from both the high-density EMG (SIL = 0.91) and the high-density MMG (SIL = 0.95).
However, while the spike train is perfectly reconstructed from the high-density MMG, the rate-of-agreement is only \SI{93}{\%} for the EMG-based decomposition.
The last selected motor unit can only be reliably decomposed from the high-density MMG.
The poor reconstruction of the spike train estimated from the high-density EMG is both reflected in the silhouette coefficient ($\text{SIL} = 0.78$) and in the rate-of-agreement ($\text{RoA} = \SI{5}{\%}$).\\
A summary of all conducted \textit{in silico} trials is provided in Figure~\ref{fig:roa_fat}.
Therefore, the fraction of separable motor units, i.e., the number of motor units with $\text{SIL}^k > 0.9$ divided by the number of active motor units, is correlated with the thickness of the subcutaneous fat and the contraction intensity.
It can be observed that for both the high-density EMG and the high-density MMG the fraction of separable motor units decreases with the thickness of the subcutaneous fat tissue layer.
Further, the fraction of separable motor units is negatively correlated with the contraction intensity.
Notably, for all simulated conditions the decomposition performance of the high-density MMG is superior over the decomposition performance of the high-density EMG.
In detail, the fraction of separable motor units increases between \SI{56}{\%} and \SI{116}{\%} when decomposing high-density MMG instead of the high-density EMG.
The robustness of the spike train estimation is reflected in the rate-of-agreement.
That is, the mean rate-of-agreement for the motor units that are classified as reliable decomposable is \SI{98.1}{\%} for the high-density EMG and \SI{99.7}{\%} for the high-density MMG.
The mean rate-of-agreement for the motor units that are classified as not reliably identifiable, i.\,e., $\text{SIL}^k < 0.9$, is \SI{20.5}{\%} for the high-density EMG and \SI{64.3}{\%} for the high-density MMG.
This indicates that the chosen uncertainty measure might underestimate the potential performance benefit when decomposing high-density MMG instead of high-density EMG.

% ----------------------------------------------- % 
\subsection{Difference of EMG-based and MMG-based decompositions}\label{sec:res_diff_separable_mus}
\begin{figure*}
    \centering
    \includegraphics[width=1\textwidth]{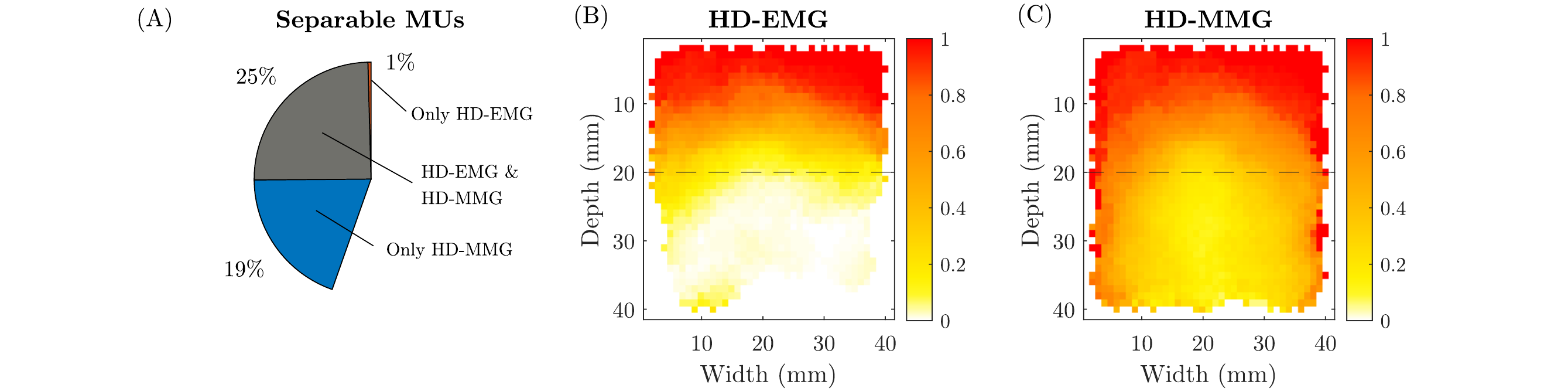}
    \caption{
        (A) Fraction of motor units that can be reliably decomposed in all \textit{in silico} trials. 
        (B) Fraction of motor units identifiable with HD-EMG depending on the spatial position of the motor units.
        (C) Fraction of motor units identifiable with HD-MMG depending on the spatial position of the motor units.
        The dashed horizontal lines in (B) and (C) show the boundary between the superficial and the deep parts of the muscle.
    }
    \label{fig:depth_comparison}
\end{figure*}

% ----------------------------------------------- %
In the previous section, it was shown that more motor units can be decomposed from high-density MMG than from the corresponding high-density surface EMG.
This section investigates if there is a difference between motor units, which can be decomposed from MMG and EMG.
Figure~\ref{fig:depth_comparison}A subdivides the motor units that are classified as reliably decomposable into three groups.
That is, \SI{25}{\%} of the motor units can be reliably decomposed from both high-density EMG and high-density MMG, \SI{1}{\%} of the motor units are only identifiable from high-density EMG and \SI{19}{\%} of the motor units can only be reconstructed from high-density MMG.
In summary, non-invasive MMG-based motor unit decomposition nearly identifies all motor units that can be decomposed from high-density surface EMG.
However, the MMG decomposition allows to observe motor units that cannot be identified with the EMG-based decomposition.\\
Next, we want to study if the motor units that can only be detected through high-density MMG-based decompositions have common characteristics.
Figure~\ref{fig:depth_comparison}B and Figure~\ref{fig:depth_comparison}C map for both EMG and MMG the fraction of identifiable motor units to the anatomical position of the motor units. 
It can be observed that the EMG-based decomposition has a strong bias to detect superficial motor units. 
In detail, when the muscle is subdivided into a superficial part and a deep part, \SI{67}{\%} of the motor units from the superficial part are classified as separable.
However, only \SI{4}{\%} of the motor units from the deep part are classified as detectable.
The MMG-based motor unit decomposition also shows a tendency to identify more superficial motor units, cf. Figure~\ref{fig:depth_comparison}C.
However, considerably more deep motor units can be reliably decomposed, i.e., \SI{88}{\%} of the superficial motor units and \SI{41}{\%} of the deep motor units are classified as detectable.

% ----------------------------------------------- % 
\subsection{Relevance of MMG properties}\label{sec:results_relevance_of_mmg_comp}
\begin{figure}[h!]
    \centering
    \includegraphics[width=0.5\textwidth]{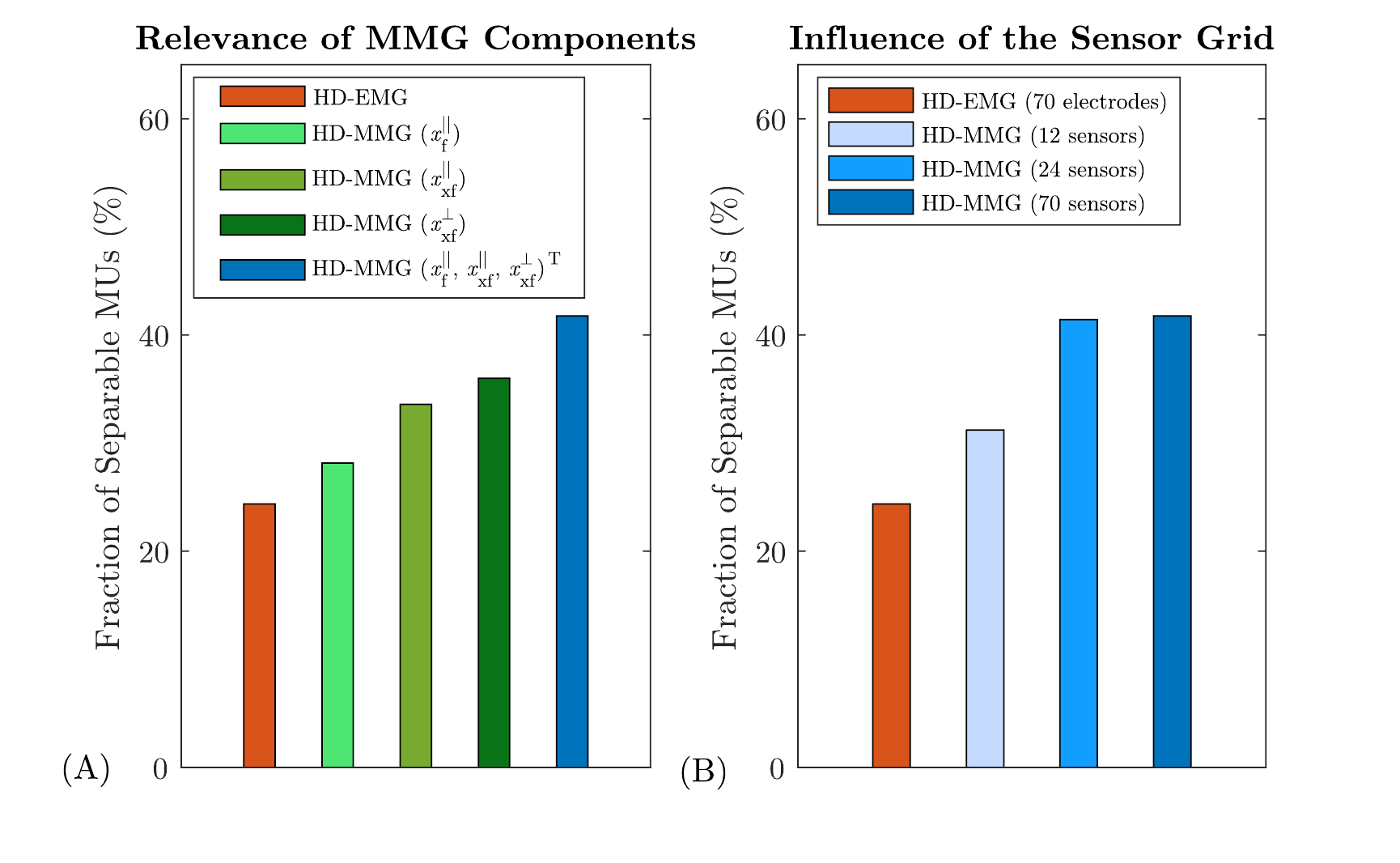}
    \caption{
    (A) Decomposition performance of the individual (scalar) MMG components.
    (B) Decomposition performance depending on the density of the MMG array.
    }
    \label{fig:comparison_fraction_of_decomposed_mus_from_sMMG}
\end{figure}

% ----------------------------------------------- % 
Next, we investigate the importance of different MMG properties, i.e., the different vector field components and the density of the sensor array, to achieve optimised decomposition results.
First, the relevance of the individual MMG components is examined.
Hence, the proposed \textit{in silico} motor unit decomposition testing framework (cf. Section~\ref{sec:methods_pipeline}) is applied to each MMG component individually. 
Notably, there are infinite possibilities for the orientation of the sensor coordinate system.
Inspired by the structure of the muscle, we consider one component aligned with the muscle fibres $x_\mathrm{f}^{\parallel}$, one component transversal to the muscle fibres as well as tangential to the body surface $x_\mathrm{xf}^{\parallel}$ and one component transversal to the muscle fibres as well as normal to the body surface $x_\mathrm{xf}^{\perp}$ (cf. Figure~\ref{fig:graphical_abstract}F).
It can be observed from Figure~\ref{fig:comparison_fraction_of_decomposed_mus_from_sMMG}A that the decomposition performance of the individual MMG components is between the goodness of the EMG-based decomposition and the MMG-based decomposition considering all magnetic field components.
Notably, the MMG components transversal to the muscle fibre direction yield a better decomposition performance than the MMG component aligned with the muscle fibres .
Thereby, the fraction of separable motor units is highest for the MMG component normal to the body surface.
In detail, defining the vector MMG-based decomposition as reference solution, the fraction of separable motor units decreases by \SI{34.4}{\%}, \SI{24.4}{\%} and \SI{18.0}{\%}, for the scalar MMG components $x_\mathrm{f}^{\parallel}$, $x_\mathrm{xf}^{\parallel}$ and $x_\mathrm{xf}^{\perp}$, respectively.\\
Next, we study the influence of the density of the MMG sensors. 
For this purpose, the signal processing workflow is applied to two additional virtual MMG data sets.
One that only considers every third MMG sensor and another one that considers every sixth sensor.
The corresponding decomposition results are summarised in Figure~\ref{fig:comparison_fraction_of_decomposed_mus_from_sMMG}B.
It can be observed that with only 12 vector magnetometers, the high-density MMG decomposition identifies more motor units than the EMG-based decomposition, which uses an array of 70 electrodes.
Further, for an array with 24 vector magnetometers, the decomposition performance is nearly as good as for the 70-MMG-sensors-case.
In detail, using only every third and every sixth MMG sensor, decreases the fraction of separable motor units by \SI{2.6}{\%} and \SI{34.2}{\%}, respectively.

% ----------------------------------------------- % 
\section{Discussion}\label{sec:discussion}
% ----------------------------------------------- % 
We analysed the performance of motor unit decomposition based on non-invasive high-density MMG and high-density surface EMG.
The latter is considered to be the current gold standard to identify the activity of individual motor units non-invasively and \textit{in vivo} \citep[e.g.,][]{DeLuca2006,Negro2016,Holobar2010,Chen2015}.
Here, we demonstrate that the decomposition of high-density MMG is superior over high-density surface EMG-based motor unit decomposition.
We rate, non-invasive MMG-based motor unit decomposition as a promising alternative to the well-established surface EMG-based motor unit decomposition.\\

\subsection{Insights on muscle-induced bio-electromagnetic fields}
The results presented in this work are predictions from a biophysically realistic computer model.
The use of an \textit{in silico} testing framework has several advantages:
first, we note that the decomposition of experimentally recorded EMG or MMG signals requires to solve a blind source separation problem.
Thereby, the decomposition results are influenced by the selected algorithms.
In the \textit{in silico} environment we can always achieve the best achievable decomposition.
This is possible as we can use the information about the forward model to solve the inverse problem.
Yet, as the motor unit decomposition problem is typically ill-conditioned, even with the full knowledge of the forward model a perfect reconstruction of the spike trains is not always feasible.\\
Further, during \textit{in vivo} experiments the biophysical properties of the body are associated with considerable levels of uncertainty. This makes it very challenging to relate the results of motor unit decompositions and the properties, the anatomy and the function of the body.
Within this work, we show that the decomposition goodness for both EMG and MMG is negatively correlated with the thickness of subcutaneous fat tissue as well as the contraction intensity.
This is consistent with decomposition results from experimentally measured surface EMG signals \citep[e.g.,][]{Farina2008b,DelVechio2020,DeOliveira2022}, even if other factors may play a role \citep{Taylor2022}.\\
The simulation environment also allows to systematically study the differences between EMG-based and MMG-based motor unit decomposition.
Notably, the decomposition of high-density MMG detects nearly all motor units that can be decomposed from the corresponding high-density EMG signal.
However, while surface EMG-based decomposition is limited to detect motor units up to a depth of approximately \SI{20}{\milli\meter}, the non-invasive MMG-based decomposition can also detect deeper motor units.
The predicted depth limit of the surface EMG-based motor unit decomposition is perfectly in agreement with experimental observations \citep[cf.][]{Fuglevand1992,Roeleveld1997b}. 
This underlines the predictive power of the presented \textit{in silico} trials.\\
Finally, this work explores which properties of a MMG recording are particularly relevant to achieve good decomposition results.
Typically, the fact that the magnetic permeability in the human body and in free space is identical is considered as the major advantage of magnetographic recordings \citep[e.g.,][]{Klotz2022}.
In this work we exemplary show that surface EMG has a more pronounced signal decay than non-invasive MMG (cf. Figure~\ref{fig:basic_signal_properties}C). This is mainly caused by the conductivity jump at the muscle boundary, i.e., limiting the amount of current that can flow transversal to the muscle fibre direction \citep[e.g.,][]{Lowery2002}.
This can explain the observation that measuring a single component of the magnetic field vector yields better decomposition results than the surface EMG-based motor unit decomposition.
Yet, this work shows that the most important requirement for MMG detection systems optimised for motor unit decomposition is the use of vector magnetometers.
This is demonstrated by the fact that even when the density of the MMG array is reduced, the vector space decomposition is superior over the EMG decomposition as well as the decomposition of the scalar MMG components. It is expected that this is caused by the unique relation between the anatomy of the motor unit territory and the spatial patterns of the MUMFs (cf. Figure~\ref{fig:basic_signal_properties}A). Accordingly, the uniqueness of the the MUMFs shows no considerable correlation with the motor unit depth and the fat tissue thickness. Yet, the MMG's amplitude decay, still limits the number of deep motor units that can be reliably decomposed.\\ 

\subsection{Limitations}
The simulation results presented within this work are upper-bound accuracy estimates of motor unit decompositions.
Whether the theoretical advantages of MMG-based motor unit decomposition also apply to experimentally measured signals depends on many factors.
First, it is noted that within this work we assumed ideal magnetometers.
Yet, magnetometers that are currently used to measure biomagnetic fields, most importantly, the superconducting quantum interference device (SQUID) \citep[e.g.,][]{Korber2016,Clarke2018} and optically pumped magnetometers (OPMs) \cite[e.g,][]{Boto2017,Osborne2018,Sander2020,gutteling2023new}, still have some limitations for high-density MMG measurements. For example, SQUIDs require cryogenic cooling and hence, the sensors must be placed several centimetres away from the skin. Further, SQUIDs are rigid devices (typically in a helmet-like geometry) that do not provide the flexibility that is required to cover different muscles. OPMs can overcome those limitations, however, still have shortcomings regarding the bandwidth (most sensors act as a low-pass filter with a cutoff frequency in the range of \SIrange{150}{300}{Hz}) and the achievable grid-density (typical sensor footprint sizes are in the range of \SIrange{1}{4}{cm^2}). The application of other magnetometers to measure MMG is currently subject of basic research, for example, magnetoelectric sensors \citep{zuo2020ultrasensitive} or nitrogen-vacancy centres \citep{zhang2022optimizing}.
Further, we assumed that the signals are not affected by noise.
However, experimentally measured EMG or MMG signals are affected by different sources of physiological noise, e.g., cross-talk from other muscles or (small) motions, and non-physiological noise, e.g., ambient electromagnetic fields or imperfections of the detection system.
The influence of these factors needs to be examined in the future.
We also note that solving a blind source separation problem strongly depends on the implemented optimisation schemes.
Moreover, the utilised biophysical model is an idealised representation of the underlying physiology and anatomy.
For example, the motor units are characterised by the mean firing rate, the muscle fibre diameter, the territory and their fibre load.
However, the distribution of the respective parameters and the validity of the corresponding assumptions might vary considerably between different muscles, subjects or patients.
This shortcoming is substantiated by the fact that we considered an idealised cube-shaped tissue geometry.
Despite the advantage that geometrical effects do not play a role in the presented results, the influence of the muscle geometry needs to be explored in the future. 
This can easily achieved by applying the proposed methodology to realistic muscle geometries discretised by the finite element method \citep[cf. e.g.,][]{Mordhorst2015,Schmid2019,Lowery2002}.\\

\subsection{Conclusion and Outlook}
In conclusion, this work shows that high-density MMG has the potential to become the new gold standard for recording single motor unit activity non-invasively and \textit{in vivo}.
Further experimental research is required to proof the theoretical predictions presented within this work. 
It is anticipated that this will be feasible within the next decade.
Thereby, the proposed \textit{in silico} trial framework can be used to optimise novel high-density MMG recording systems, assist the interpretation of experimental measurements and benchmark the performance of motor unit decomposition algorithms.

\section*{Acknowledgements}
This research was funded by the Deutsche Forschungsgemeinschaft (DFG, German Research Foundation) under Germany's excellence strategy (EXC 2075 390740016) and the European Research Council (ERC) through the ERC-AdG ``qMOTION'' (Grant agreement ID: 101055186; O.R.) as well as the ERC-CoG ``INcEPTION'' (Grant agreement ID: 101045605; F.N.). 

\def\newblock{\ }%
\bibliographystyle{abbrvnat}      % basic style, author-year citations
\bibliography{references}

\end{document}